\documentclass[letterpaper,twocolumn,10pt]{article}
\usepackage{usenix,epsfig,endnotes,hyperref,paralist}
\begin{document}
\thispagestyle{empty}
\DeclareGraphicsExtensions{.png}
\date{}
\title{\Large \bf Cryptocat: Adopting Accessibility and Ease of Use as Security Properties}

% \author{
% {\rm Author A}\\
% \and
% {\rm Author B}\\
% }

\author{
{\rm Nadim Kobeissi}\\
\href{https://crypto.cat}{Cryptocat}, \href{http://openitp.org/}{OpenITP}\\
\href{mailto:nadim@crypto.cat}{nadim@crypto.cat}
\and
{\rm Arlo Breault}\\
\href{https://crypto.cat}{Cryptocat}\\
\href{mailto:arlo@crypto.cat}{arlo@crypto.cat}
}

\maketitle

\section*{Abstract}

Cryptocat is a Free and Open Source Software (FL/OSS) browser extension that makes use of web technologies in order to provide easy to use, accessible, encrypted instant messaging to the general public. \\
We aim to investigate how to best leverage the accessibility and portability offered by web technologies in order to allow encrypted instant messaging an opportunity to better permeate on a social level. We have found that encrypted communications, while in many cases technically well-implemented, suffer from a lack of usage due to their being unappealing and inaccessible to the ``average end-user". \\
Our position is that accessibility and ease of use must be treated as security properties. Even if a cryptographic system is technically highly qualified, securing user privacy is not achieved without addressing the problem of accessibility. Our goal is to investigate the feasibility of implementing cryptographic systems in highly accessible mediums, and to address the technical and social challenges of making encrypted instant messaging accessible and portable.

\section{Introduction}

Current popular encrypted instant messaging technologies largely implement the Off-the-Record protocol (OTR) \cite{otr} for encryption between two parties. OTR aims to provide forward secrecy, digital signatures, message authentication, repudiation and plausible deniability for conversations with two participants. OTR-encrypted chat is generally available as a plugin for popular instant messaging software \cite{otr-pidgin,otr-adium,otr-jitsi,otr-gajim}. For mobile smartphones, OTR is available built-into specialized encrypted messaging applications \cite{gibberbot,chatsecure}. \\
We have found that the mobile applications featuring OTR tend to be more accessible due to their platform, specialized purpose and design philosophy which integrates OTR from the outset. However, in the case of desktop applications, we have found that the necessity for both parties to download, install and configure the same chat software and OTR plugin was enough to disenfranchise a majority of end-users from regularly engaging in encrypted messaging, even if they had an urgent or pressing need for encrypted communications. \\
In working with young and middle-aged professionals in the Middle East region, we have discovered that desktop OTR clients suffer from serious usability issues which are sometimes further exacerbated due to language differences and lack of cultural integration (the technology was frequently described as ``foreign"). In one case, an activist who was fully trained to use Pidgin-OTR \cite{otr-pidgin} neglected to do so citing usability difficulties, and as a direct consequence encountered a life-threatening situation at the hands of a national military in the Middle East and North Africa region (see \S7.1). \\
These circumstances have led us to the conclusion that ease of use and accessibility must be treated as security properties, since their absence results in security compromises with consequences similar to the ones experienced due to cryptographic breaks. \\
Cryptocat is designed to leverage highly accessible mediums (the web browser) in order to offer an easy to use encrypted instant messaging interface accessible indiscriminately to all cultures, languages and age groups. Cryptocat clients are available as Free Software browser extensions \cite{cryptocat} written in JavaScript and HTML5. Cryptocat servers use the XMPP protocol \cite{xmpp} with XEP-0045 \cite{xep-0045}.

\section{Goals}

Cryptocat's goal is to profoundly broaden the accessibility of encrypted chat across platforms, cultures, languages and borders. To achieve this, we aim to expand the limits of cryptosystem implementation in highly accessible environments. \\
Our main target is the web browser: a best-of-breed environment in terms of accessibility but with a deeply lacking research base regarding its handling of client-side cryptographic systems. We aim to improve the research base and legitimacy of secure cryptography in the browser in order to allow for the implementation of highly accessible and portable cryptosystems. Technical experimentation is required, as well as directly addressing issues such as code delivery and secure pseudo-random number generation (see \S6). \\
On a social level, Cryptocat aims to normalize the expectation that instant messaging conversations must be encrypted against any and all undesired third parties. If we can achieve a high level of accessibility for encrypted instant messaging platforms, we hope that our methodologies will be widely adopted through our Free and Open Source development model, thereby normalizing the idea of private online conversation and embedding a social expectation for communication privacy. Cryptocat also aims to promote cats around the world as wonderful creatures.

\section{Threat model}

Cryptocat's security objectives are:
\vspace{0mm}
\begin{itemize}
\setlength{\itemsep}{1pt}
\setlength{\parskip}{1pt}
\item Provide encrypted messaging where messages are only readable by the sender and the intended recipient(s).
\item Provide means for parties to securely authenticate each other's identities.
\item Protect against message forgery during conversations.
\end{itemize}
\begin{figure}[t]
\centering
\includegraphics[scale=0.40]{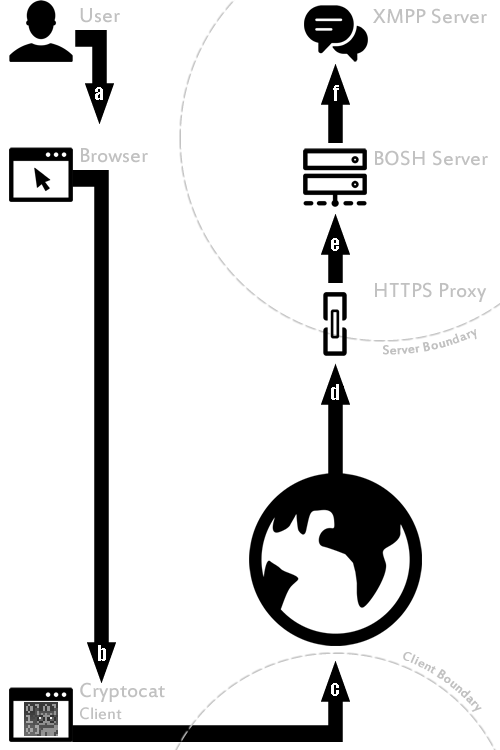}
\caption{Connections overview and attack points}
\end{figure}
It is worth noting that our security objectives do not include protecting from compromise via hardware or malware, or the cryptographic property of plausible deniability in a group conversation context, or anonymizing the identity of user connections. That being said, plausible deniability is still provided in private messages since these are carried over the OTR protocol. Futhermore, Cryptocat clients are tested to be compatible with third-party anonymizing technologies such as Tor \cite{tor}, and it is possible for Cryptocat servers to be set up as Tor Hidden Services \cite{tor-rendezvous}.  \\
Cryptocat's threat model document \cite{threatmodel} delineates six attack points with potential threats, pinpointed along the route from the Cryptocat user to the server. Threats are evaluated according to the DREAD model \cite{dread} (Damage, Reproducibility, Exploitability, Affected users and Discoverability). The attack points (A, B, C, D, E and F) are visible in Fig. 1.\\
The full threat model document \cite{threatmodel} discusses a variety of threats situated along these attack points, which range from a phishing browser extension masquerading as a Cryptocat application to SSL man-in-the-middle attacks \cite{schneier-ssl}. Our threat evaluation process is continuous and open to public discussion.

\section{Methodology}

The reasoning behind our development methodology is that Cryptocat is most likely to receive volunteer programmers, testers and security auditors via this transparent format. We believe in principles of \emph{full disclosure} and aim to achieve our security and usability through the detailed and open study of our protocols, implementations and releases, and by maximizing the involvement of both security and usability communities.

\subsection{Software development}

Cryptocat software is written, published and reviewed under a principle of \emph{radical transparency}. All source code modifications are pushed live into a public code repository using the git revision control and source code management system, which also hosts an issue tracker and development wiki. All design considerations, from cryptographic implementations to user interface concerns, are discussed, decided upon and implemented in public. Some larger projects may be developed offline prior to the first commit, but no release is ever made without it being available as Free, Open Source Software and properly documented on our code repository \cite{cryptocat-code}. \\
Cryptocat browser extensions, currently available for Google Chrome, Mozilla Firefox and Apple Safari, rely on an identical codebase with only minor differences to account for browser integration. Browser extensions are signed and delivered over HTTPS. \\
Furthermore, routine compatibility and functionality testing is carried out between different browsers and platforms, in-house, via unit testing, and in collaboration with beta testers around the world.

\subsection{Auditing}

Cryptocat browser extensions undergo regular professional security audits carried out by various independent third party code auditing groups (see \S10). Unless we are forbidden by the auditor, all audit reports are made fully public as soon as any critical flaws they may identify are addressed. After our first audit \cite{audit-cure53} in November 2012, we addressed thirteen vulnerabilities, of which two were considered critical, and verified the fixes with the auditing group within one week. The audit report was made available to the public immediately after. Our most recent audit \cite{audit-veracode} was carried out by Veracode in January 2013 and discovered no vulnerabilities within its scope, awarding Cryptocat a score of 100/100. As of writing, a third code audit is underway.

\subsection{Localization}

In order to fulfill Cryptocat's accessibility goals, Cryptocat's interface is fully translated into more than 32 languages and functions well with right-to-left scripts as well as exotic character sets necessary for supporting languages such as Tibetan. \\
We use the online Transifex \cite{transifex} platform in order to coordinate with volunteer translators and to review and verify translations. We are entirely dependent on a worldwide community of volunteer translators in order to help Cryptocat break down linguistic and cultural barriers. \\
Currently maintained languages and dialects are Arabic, Basque, Bengali, Bulgarian, Burmese, Catalan, Chinese, Simplified Chinese, Croatian, Czech, Danish, Dutch, English, Estonian, Finnish, French, German, Greek, Hebrew, Italian, Japanese, Khmer, Korean, Latvian, Lolcat, Norwegian, Persian, Polish, Portuguese, Russian, Serbian, Slovak, Spanish, Swedish, Tibetan, Turkish, Uighur, Urdu and Vietnamese.

\section{Architecture}

Cryptocat's technical architecture is similar to typical XMPP chat clients with Multi-User Chat support, with some additional features and changes required to accomodate for our reliance on the web browser as a platform. Essentially, Cryptocat's architecture is split into an XMPP client capable of encrypted communications and an XMPP server accessible via HTTPS.

\subsection{Client-side}

Cryptocat is delivered as a signed browser extension, downloaded over HTTPS. The extension contains an XMPP client that only accepts and sends encrypted messages while discarding all other communications. All cryptographic operations take place on the client-side, with the server only dealing with the exchange of ciphertext messages and user login. Cryptocat is designed around a ``chat room" model where users adopt discardable nicknames to join ephemeral chatrooms, encrypting messages using ephemeral public keys. \\
Cryptocat uses different encryption protocols depending on which type of communication users are engaged in. For one-on-one chats, a JavaScript implementation \cite{otrjs} of Off-the-Record messaging \cite{otr} is used. In multiparty situations, Cryptocat currently uses its own temporary multiparty protocol \cite{cryptocat-spec}, with the intention of replacing the protocol once the Multi-Party Off-the-Record messaging protocol (mpOTR) \cite{mpotr} is specificied and implemented (see \S9.1).

\subsection{Server-side}

Connections between Cryptocat clients and servers are illustrated in Fig. 1. On the server side, Cryptocat uses ejabberd \cite{ejabberd} in order to instantiate an XMPP server. Since the Cryptocat client is built using web technologies, a Bidirectional-stream Over Synchronous HTTP (BOSH) \cite{bosh} server is used in order to instantiate an HTTP-based transport for ``push" and ``pull" XMPP operations. This HTTP transport is not exposed to the public Internet and is only mapped to the local IP address; instead, we use a separate HTTPS proxy to expose the BOSH server globally. This is done both in order to benefit from an additional layer of encryption via SSL and to be able to offer XMPP over BOSH over a standard 443 port, therefore helping mitigate potential incompatibilities with user firewalls. \\
Cryptocat servers use XEP-0045 \cite{xep-0045} to instantiate group ``chat rooms", while XEP-0077 \cite{xep-0077} is used to allow for spontaneous In-Band Registration of ephemeral user identities. \\
Server configuration files and deployment instructions are available online on the Cryptocat Development Wiki \cite{server-instructions}. Users are encouraged to set up their own private servers, and the Cryptocat client offers an interface for configuring third party server connections.

\section{Technical challenges}

Adopting the web browser as our platform has the most payoff in terms of securing accessibility and ease of use, but is also recognized as a substantially risky technical challenge. As of the introduction of the HTML5 framework \cite{html5}, it has been possible on a technical level to implement the myriad features required for a proper encrypted instant messaging client natively in the browser. However, we were faced with a deep lack of research, testing, specification and implementation when it came to certain security considerations. \\ Therefore, in many situations, we have had to come up with potential solutions to these security questions ourselves. Aside from the cases outlined in the following subsections, Cryptocat is also participating as a test case for the W3C Web Cryptography Working Group \cite{webcrypto-group}, which is currently attempting to implement native cryptographic primitives to be accessed in the browser via an API that is exposed through JavaScript calls \cite{webcrypto-spec}. We ultimately believe that even though further research is still required, it is absolutely necessary for web technologies to achieve a security standard sufficient for the implementation of sophisticated cryptographic systems, if we are to succeed in making these systems accessible to the general public.

\subsection{Code delivery}

Early attempts to deliver Cryptocat as a website over HTTPS without the requirement of installing a browser extension were deemed an exigent security risk. This is due to the fact that as a website, cryptographic primitives were re-downloaded every time a user would access the Cryptocat client. This left an open window for a malicious host to serve deceptive or defective encryption software, masquerading as a legitimate Cryptocat client. Instead, Cryptocat was rebuilt to be served as a code-signed browser extension which connects to standard XMPP servers. This made code delivery more secure and also allowed us to rely on existing standards for server deployment.

\subsection{OTR in the browser}

\begin{figure}[t]
\centering
\includegraphics[scale=0.25]{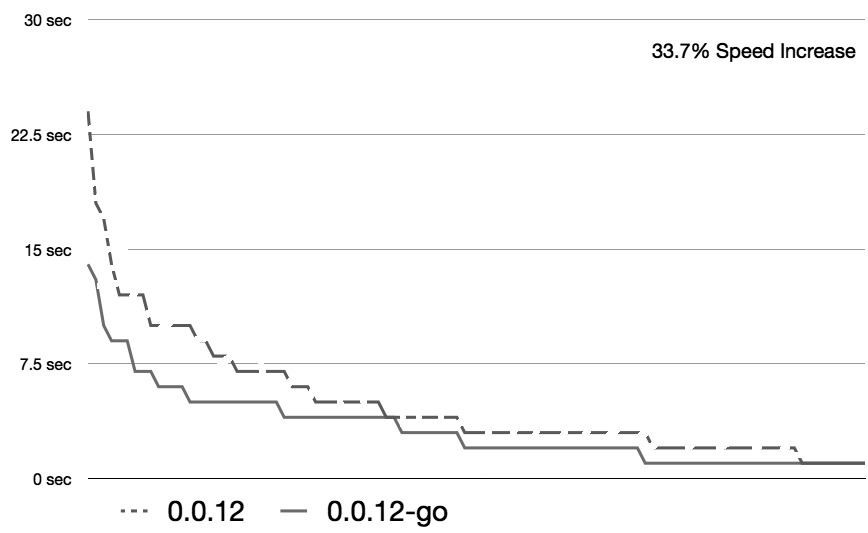}
\caption{OTR key generation speed optimization}
\end{figure}
JavaScript does not natively support arbitrary precision arithmetic. A BigInt library, along with several custom written operations, made up for this deficit using a combination of JavaScript Array and Number types. Until recently, JavaScript did not have an answer for binary data manipulation in all browsers. Typed Array support started to land in major web browsers in late 2012, well after the initial development of the library had begun. Consequently, in our OTR library, raw data is treated as JavaScript Strings, which is slow but compatible. \\
OTR uses 1024-bit DSA keys which, due to the fact that Cryptocat does not currently store long-term key pairs (see \S9.2), need to be generated, along with DSA parameters, each time the application is launched. Efficiently generating these keys is paramount to user experience. Since JavaScript's execution model is a single-threaded, run to completion event loop, computationally heavy operations will by default lock up the browser tab, preventing other operations from executing. To mitigate this issue, keys are generated in Web Workers \cite{webworkers} and message-passed back to the main thread upon completion. This allows Cryptocat a first line of defense: displaying randomly selected ``cat facts" to keep the user engaged as it generates keys. As of writing, Cryptocat features over fifty highly interesting cat facts \cite{catfacts}. \\
A large gain in performance came from an optimization to the Miller-Rabin primality test by first checking if the candidate prime is divisible by any prime number under 1,000. We further optimized the speed of OTR key generation in the browser by 33.7\% by following Google's Go-lang implementation \cite{go-dsa} of the Digital Signature Algorithm (DSA), which deviates from the FIPS 186-3 \cite{dss} standard by eliminating the verification seed used to generate primes, since it ends up being discarded without use. This significantly sped up key generation, as seen in Fig. 2. It also led to less variance in generation time. Furthermore, improvements to JavaScript virtual machines have lead to optimizing compilers that execute cryptographic operations with near-native performance \cite{v8}. These improvements make DSA key generation feasible in the browser.

\subsection{CSPRNG in the browser}

In JavaScript, the default pseudorandom number generator is not considered cryptographically secure (except in Opera, which implements a cryptographically secure version of Math.random()). \\
As of writing, Google Chrome and Apple Safari are the only web browsers to expose a Cryptographically Secure Pseudorandom Number Generator (CSPRNG) via JavaScript, using the window.crypto.getRandomValues() \cite{webcrypto-spec} function. In Mozilla Firefox, the Cryptocat browser extension manually exposes this method using a JavaScript module which implements window.crypto.getRandomValues() on a native browser level via calls to NSSLib. We have received reassurances from Mozilla that Firefox 22 will natively include window.crypto.getRandomValues(). \\
In order to minimize draining of the system entropy pool, window.crypto.getRandomValues() is only called for an initial batch of pseudorandom values, which are then used to seed a Salsa20 \cite{salsa20} implementation.

\section{Social challenges}

Challenges in socially normalizing cryptography are largely due to cryptographic applications being almost always available as sideline alternatives. Designers of mainstream instant messaging applications are rarely given the incentive to implement client-side cryptography, while the designers of encrypted communication systems usually cater to specific audiences, such as computer hobbyists and enthusiasts, military and intelligence personnel, and so on. Cryptocat has enjoyed a highly varied user-base, ranging from LGBTQ help groups to pre-school teachers using it to teach children about online privacy. \\
In order for encrypted instant messaging to become a new mainstream standard, we are required first to develop the Cryptocat browser extension as an accessible, easy and fun to use software package, while also ensuring that the underlying research is available to be independently implemented into the clients of other, more mainstream instant messaging solutions. \\
Cryptocat aims to leverage the web browser environment in order to make itself readily accessible to users who are already familiar with mainstream instant messaging solutons that use the same environment, such as Facebook Chat. For this purpose, Cryptocat uses a neutral interface with comforting color palettes and graphics, coupled with usability features such as audio and desktop notifications. We classify these interface features also as security features, since they encourage users to engage in a more secure form of communication, especially when it is needed. We also solicit the aid of volunteer translators in order to maintain Cryptocat translations that help us lessen the effect of linguistic and cultural barriers.

\subsection{Unintended demographics}

We have found that in attempting to standardize encrypted instant messaging as accessible to everyone, there is a serious danger of being misinterpreted as catering specifically to political activists in oppressive regimes and others in similarly life-threatening situations. While it is understandable as to why these groups would gravitate around a relatively highly accessible encrypted instant messaging solution, we have found that clearly delineating Cryptocat's efforts as a much broader experimental nature to be a demanding and serious task. As of writing, warnings and tips regarding best-use practices, translated into thirty-five languages, are included on the Cryptocat client's main screen. \\
Regarding the issue of addressing the privacy needs of those in dangerous situations, we deem it necessary to never rely exclusively on software, but to focus on providing capable training sessions regarding the risks involved.

\subsection{Buddy lists}

Discussions were held early in Cryptocat's development to decide on whether features such as user accounts and buddy lists were to be implemented. While such features would greatly enhance user experience and ease of use by allowing users to quickly communicate with friends and check on the status of their contacts, we have decided against implementing these features in order to minimize the amount of user metadata held on Cryptocat XMPP servers. Were these features implemented, the Cryptocat XMPP server would have likely recorded, in plain text, the contact lists and user account information of all of its users. In order to maximize user privacy and minimize server data retention, we have therefore opted for a ``chat room" model, even though that model is arguably less user-friendly. This example illustrates the balancing act between usability and security that has provided the most central and essential questions surrounding Cryptocat's development decisions.

\subsection{Encouraging authentication}

We have found that average users generally do not bother authenticating each other's identities using public key fingerprints, due to the task's cumbersome nature. In order to facilitate this, we have introduced public-key derived color codes that provide a limited, albeit more likely to be used, capacity to verify identities. We are also implementing the Socialist Millionaires' Problem protocol in order to allow for in-band question/answer authentication (see \S9.4).

\section{Availability and status}

Cryptocat is currently available as a browser extension for Google Chrome, Mozilla Firefox and Apple Safari through the Cryptocat website \cite{cryptocat}. It is currently in Beta, experimental status.

\section{Limitations and future work}

Although browser tab thread-level sandboxing is quickly improving and increasingly more difficult to break, especially in the case of Google Chrome \cite{chrome-sandboxing}, we believe that the browser's architecture, which is designed to handle and execute myriad amounts of different data types, needs additional research in the area of sandboxing. Improvements in this area will greatly aid the web browser with securely handling cryptographic data. We also believe that research is required towards implementing other security features in JavaScript, such as secure memory erasure for cryptographic keys.

\subsection{Implementing mpOTR}

The lack of a Multi-Party Off the Record protocol specification and implementation remains a serious impediment towards achieving more secure encrypted group instant messaging. While research material currently exists regarding mpOTR \cite{mpotr}, specification and implementation still require serious effort due to the nature of the problem of efficiently implementing deniable multi-party encrypted chat. The current pragmatic value of plausible deniability in the OTR context is uncertain.

\subsection{Permanent key storage}

A suitable storage mechanism for long-term keys has yet to be decided upon, nor has importing long-term keys been implemented in Cryptocat. This means that DSA keys are regenerated every time the application is launched, requiring chat participants to verify fingerprints in an out-of-band channel at each conversation, or use the Socialist Millionaires' Problem protocol to verify a shared secret. Key storage and SMP are currently available in our OTR library but are still undergoing testing and review \cite{smp}.

\subsection{File transfer}

Cryptocat implements an OTR-XMPP file transfer specification \cite{filetransfer-spec} that leverages the extra symmetric keys made available in the third version of the OTR messaging protocol. \\
There were three aspects to achieving secure file transfers: establishing a shared secret key between clients, choosing the cryptographic primitives used for file encryption, and selecting a transfer protocol. In the third version of the OTR messaging protocol, an extra symmetric key is derived during authenticated key exchanges, with the intention of being used for secure communication over a different channel. Therefore, an API is provided in the JavaScript OTR library, and used in Cryptocat, to derive a shared 256 bit key. We then run this shared key through SHA512 to expand it to 512 bits. The first 256 bits of the resulting expanded key are used with AES-256, in counter mode, to encrypt chunks of the file. The second 256 bits are used in HMAC-SHA512 for authentication. To transfer the file, we implemented the XEP-0047 specification \cite{ibb}, which defines an XMPP protocol to enable entities to establish a bytestream. Data is broken down into smaller chunks (64Kb) and transported in-band over XMPP.

\subsection{SMP}

The OTR specification provides two methods of detecting impersonation or man-in-the-middle attacks: comparing fingerprints and the Social Millionaires' Protocol (SMP). Due to the current limitation involving long-term key storage, verification of fingerprints in a different channel at each conversation is required. SMP offers a more convenient alternative and, as we have argued, ease of use will encourage clients to make use of this security feature. SMP allows two users to compare a secret without revealing any information about that secret, other than whether or not it is known.

\section{Acknowledgments}

We owe many advancements in Cryptocat's research and development to the contributions, review, testing and feedback of Jacob Appelbaum, Joseph Bonneau, Griffin Boyce, Dmitry Chestnykh, Daniel ``koolfy" Faucon, Arturo Filast\`{o}, Adam Langley, Meredith L. Patterson, Fabio Pietrosanti and the open source security community. We are indebted to Elisabeth Gill for coordinating Cryptocat's translation effort with dozens of volunteer translators. We would like to thank David Dahl and Tom Lowenthal of Mozilla, and fellow members of the W3C Web Cryptography Working Group for their help with browser integration. We thank the teams at Cure53 and Veracode for their diligent security audits of Cryptocat code. Finally, we would like to thank the Open Technology Fund \cite{otf} and OpenITP \cite{openitp} for funding our work.

\end{document}